\documentstyle[11pt,aaspp4,flushrt]{article}  

\def\msun{{\rm M_{\odot}}}
\def\rsun{{R_{\odot}}}
\def\lta{\la}
\def\gta{\ga}
\begin{document}

\title{THE FORMATION OF LOW--MASS TRANSIENT \\ X--RAY BINARIES}

\author{A.~R.~King$^1$ and U.~Kolb$^2$}
\affil{
Astronomy Group, University of Leicester,
Leicester LE1 7RH, U.K. \\
$^1$ark@star.le.ac.uk, $^2$uck@star.le.ac.uk 
}


\begin{abstract}
We consider constraints on the formation of low--mass X--ray binaries
containing neutron stars (NLMXBs) arising from the presence of soft
X--ray transients among these systems.
For a neutron star mass $M_1 \simeq 1.4\msun$ at formation, 
we show that in short--period ($\lta 1 - 2$ d) systems 
driven by angular momentum loss
these constraints require the secondary at the beginning of 
mass transfer to have 
a mass $1.3\msun \lta M_2 \lta 1.5\msun$, and to be significantly
nuclear--evolved, provided that supernova (SN) kick velocities are
generally 
small compared with the pre--SN orbital velocity.
As a consequence a comparatively large fraction of such systems
appear as soft X--ray transients even at short periods,
as observed. Moreover the large initial
secondary masses account for the rarity of NLMXBs at periods $P \lta 3$ hr.
In contrast, NLMXB populations forming with large kick velocities would
not have these properties, suggesting that the kick
velocity is generally small compared to the pre--SN orbital
velocity in a large fraction of systems,
consistent with a recent reevaluation of pulsar proper motions.
The results place also tight constraints on the strength of magnetic
braking: if magnetic braking is significantly stronger than the standard
form too many unevolved NLMXBs would form, if it is
slower by only a factor $\simeq 4$ no short--period NLMXBs would form
at all in the absence of a kick velocity.
The narrow range for $M_2$ found for negligible kick velocity implies
restricted ranges near $4 \msun$   
for the helium star antecedent of the neutron star and
near $18\msun$ for the original main--sequence progenitor. The 
pre--common envelope period must lie  near  4 yr,
and we  
estimate the short--period NLMXB formation rate in the disc of the Galaxy as 
$\sim 2 \times 10^{-7}$ yr$^{-1}$. Our results show that the neutron
star mass at short--period
NLMXB formation cannot be significantly larger than $1.4\msun$.
Systems with formation masses $M_1\lta 1.2\msun$ would have
disrupted, so observations implying $M_1 \sim 1.4\msun$ in some NLMXBs suggest
that much of the transferred mass is lost from these systems.

\end{abstract}

                 
\newpage

\section{INTRODUCTION}
In a recent paper (King et al, 1996; hereafter Paper 1)
we considered the disc instability model
for soft X--ray transients (SXTs). Following van Paradijs (1996) we noted that
X--ray irradiation of the disc surface tends to suppress the instability,
so that SXT behaviour requires 
that the mass transfer rate is below a critical limit which is itself
a rising function of orbital period $P$.
For $P \gta 1 - 2$ d mass transfer is driven by the nuclear
expansion of an evolved secondary, and proceeds at low enough rates
that almost all of these systems appear as SXTs. For $P \lta 1 - 2$ d
mass transfer is driven by angular momentum losses; in Paper 1 we
showed that the resulting rates are higher than the critical rate, hence
too high for SXT behaviour unless the secondary star is already somewhat
nuclear--evolved before mass transfer begins. 
This effect is particularly
marked in a neutron--star low--mass X-ray binary
(neutron--star LMXB, or NLMXB): if we write (cf Paper 1) $\hat m_2$ for the
ratio of the secondary's mass to that of a ZAMS star filling the Roche lobe
at the same orbital period, SXT behaviour requires $\hat m_2 \lta 0.25$.
By contrast
we found a much weaker limit $\hat m_2 \lta 0.75$ for SXT behaviour in
a LMXB with a $10\msun$ black--hole primary. These limits agree with
the extreme mass ratios always found in SXTs. In Paper 1 we suggested that
these limits also offered a potential explanation for the prevalence of
black--hole systems among SXTs (8 out of 14 with known $P$) as compared with
persistent LMXBs (1 out of 29 with known $P$). As NLMXBs can be reliably 
identified by the presence of X--ray bursts, we consider this problem
further by investigating constraints on NLMXB formation.

\section{NLMXB FORMATION}

The requirement that short--period SXT secondaries should be significantly
evolved is a powerful clue in investigating LMXB (and particularly NLMXB)
formation. For example, the very fact that at least 4 NLMXBs
out of 22 in the relevant period range are transients 
(0748-676, Cen X-4, 1658-298, Aql X-1; cf.\ van~Paradijs 1995) 
must imply that NLMXB secondaries
usually have masses $M_2 \gta 1\msun$ at the onset of mass
transfer. Lower initial--mass secondaries would be essentially unevolved when
mass transfer began, and a large population
of them would require the 
4 known SXTs among NLMXBs to be accompanied by a far greater number of
persistent NLMXBs than observed. 
However to get any NLMXBs at all at $P \lta 1 - 2$~d
simultaneously requires that some of the secondaries initially have masses
$M_2 \lta 1.5\msun$ so that angular momentum loss via magnetic braking can
shrink the binary. 
In addition, for $M_2 \gta 1.5\msun$ these NMLXBs with a
main--sequence donor will be unstable to thermal--timescale mass
transfer once contact is reached; they transfer mass at
super--Eddington rates and do not appear as NLMXBs (Kalogera \&
Webbink 1996a). 
Given these restrictions all NLMXBs with $P \lta 1 - 2$ d 
(and indeed most at longer periods too) must apparently form with $M_2$ in
a very restricted range $1\msun \lta M_2 \lta 1.5\msun$. 

The properties and the very existence of this group depend sensitively
on the efficiency of magnetic braking.
The paper by Iben et al.\ (1995b) on general LMXB formation
effectively rules out these systems
because it assumes that magnetic braking does not operate for $M_2 >
1\msun$. Similarly Terman et al.\ (1996) do not consider them
in detail as they somewhat artificially assume that $M_2 = 1.0 \msun$ 
separates systems captured by angular momentum losses from systems
evolving towards longer orbital period.

Below we investigate in detail constraints on NLMXBs forming
after a  common envelope (CE) phase and subsequent helium star
supernova (SN). 
We consider first the fundamental case of a
spherically symmetrical SN (Sect.~2.1), and we shall show that in this
case secondary stars in NLMXBs are indeed significantly
nuclear--evolved at the onset of mass transfer. A generalization to
the more realistic case when the neutron star receives a kick velocity
during the SN (Sect.~2.2) reveals that the fraction of systems with 
unevolved secondaries increases with increasing magnitude of the
kick. For a given magnetic braking strength this places limits on the
magnitude of the kick velocity. 
Possible alternative evolutionary channels leading to NLMXB
formation are discussed in Sect.~2.3.

\subsection{NLMXBs from spherically symmetrical helium--star supernovae}

To discover any limits on the initial mass $M_2$
of the secondary we consider the constraints on the formation of
LMXBs from helium--star + main--sequence (MS)
binary remnants of common envelope evolution in the case of a
spherically symmetrical SN. 

\subsubsection{Constraints on the formation}

Webbink \& Kalogera (1994; hereafter WK) and Kalogera \& Webbink
(1996a) list a 
number of requirements that the various progenitor stages must satisfy in
order to lead to LMXB formation. Of these, three turn out to be crucial in
fixing $M_2$:

(a) The post--common--envelope binary must be wide enough to allow the helium
star to evolve to core collapse (requirement 6 of WK).

(b) The binary must survive the supernova event resulting from the core
collapse of the helium star (7 of WK).

(c) After the helium star progenitor of the LMXB primary
explodes as a supernova the binary must be able to
reach interaction within the age of the Galaxy (WK 8).

If $M_2$ is large enough and the post--SN separation is wide
enough (c) will occur through the nuclear expansion of the secondary.
Such systems require $M_2 \gta 1\msun$ 
to give main--sequence lifetimes $t_{\rm MS}$ less than 
the age $t_{\rm Gal}$ of the Galaxy and must have 
$P \gta 1 - 2$ d at the onset of mass transfer; $P$ will increase
further as the secondary expands. 
However if $M_2$ is less than this and/or the post-SN
binary is narrower, (c) can come about through orbital
angular momentum loss. After reaching contact with an initial orbital period
of order 1 -- 2 d, such systems evolve either to longer
($P \gta 1$ d) or shorter ($P \lta 1$ d) periods depending on the
competition between nuclear evolution and angular momentum loss
(Pylyser \& Savonije 1988a,b).

For the remainder of this paper we concentrate on this latter group
of systems, which are the only ones which can populate the short--period
($P \lta 1 - 2$ d) region.
It is convenient to apply the constraints (a -- c) in a different order. 
From (c) we require $t_{\rm MB}, t_{\rm GR} < t_{\rm Gal}$, where
\begin{equation}
t_{\rm MB} = 2.2\times 10^8 \, \frac{\gamma \: m_1}{{(m_1+m_2)}^{1/3}
         m_2^4} P_d^{10/3} 
         \biggl[1 - \biggl({P_{\rm c}\over P}\biggr)^{10/3}\biggr]\ {\rm yr}
\label{eq1}
\end{equation}
\begin{equation}
t_{\rm GR} = 4.7\times 10^{10} \, \frac{{(m_1+m_2)}^{1/3}}{m_1 m_2} P_d^{8/3}
         \biggl[1 - \biggl({P_{\rm c}\over P}\biggr)^{8/3}\biggr]\ {\rm yr}
\label{eq2}
\end{equation}
are the timescales for detached orbital evolution under
magnetic braking and gravitational radiation, with $\gamma$ a dimensionless
efficiency parameter for the former case. 
These shrink the post--SN binary from its
initial period $P$ to the value $P_{\rm c}$ at which it first comes
into contact with its
Roche lobe ($P_d = P/{\rm d}, m_1 = M_1/\msun, m_2 = M_2/\msun$ etc).
If the secondary is unevolved  
an adequate approximation for $P_{\rm c}$ is 
$P_{c} \simeq 0.375 \, m_2\ {\rm d}$,
while $P_{\rm c}$ is up to an factor $\simeq
2.5$ larger if the secondary is fairly massive ($M_2 \gta 1 \msun$)
and close to the terminal main--sequence.

We assume that the secondary has a structure such that a magnetic
stellar wind brakes its rotation; since we are considering
short--period LMXBs in which the
secondaries are not too far from the main sequence this requires
$0.3\msun \lta M_2 \lta 1.5\msun$. Furthermore we assume that the
secondary is tidally locked to the orbit such that magnetic
braking removes orbital (rather than only rotational) angular
momentum. Applying simple
scalings for the tidal synchronization timescale (e.g.\ Tassoul 1995)
suggests that this is the case for the detached systems in the
period range under consideration (see below).
The efficiency parameter $\gamma$ in (\ref{eq1}) allows us to test the
sensitivity of  
our results to the strength of magnetic braking. The standard case 
$\gamma = 1$ corresponds to the description by Verbunt \& Zwaan (1981)
when the radius of gyration is set to $\sqrt{0.2}$ and the calibration 
parameter to unity.  

Requirement (c) thus gives the
longest post--SN period compatible with short--period LMXB
formation. It is easy to 
show that the longest possible value of this period is given by equating
$t_{\rm MB}$ to $t_{\rm Gal}$ with $P \gg P_{\rm c}$, so that the term
in square brackets in (\ref{eq1}) is unity.
Kepler's law now gives a condition on the separation $a_{\rm post SN}$
of the tidally--circularized post--SN binary, i.e. 
\begin{equation}
 a_{\rm postSN} < 9.0 \rsun {\biggl[ \frac{t_G}{\gamma} \: 
                  \frac{(m_1+m_2)^2 m_2^4}{m_1} \biggr]}^{1/5}, 
 \label{eq20} 
\end{equation}
where $t_G = t_{\rm Gal}/10^{10}\ {\rm yr}$.

Before exploiting (\ref{eq20}) further we apply requirement (b). 
In a spherically symmetrical supernova
explosion, it is well known that the remnant binary is disrupted if more
than one--half of the binary mass is removed, i.e. unless
\begin{equation}
M_{\rm He} - M_1 < {1\over 2}(M_{\rm He} + M_2) \label{eq7}
\end{equation}
or
\begin{equation}
m_2 > m_{\rm He} - 2m_1 
\label{eq8}
\end{equation}
the binary will be disrupted. Here $m_{\rm He}$ is the helium star
mass $M_{\rm He}$ in $\msun$. Moreover, the eccentricity $e$ of the
orbit immediately after the SN is given by 
\begin{equation}
 e = \frac{m_{\rm He}-m_1}{m_1+m_2}.
 \label{eq22}
\end{equation}
Obviously condition (\ref{eq8}) is equivalent to $e<1$.
The pre--SN separation and the post--SN separation of the tidally
circularized orbit are related by (e.g.\ Verbunt 1993)
\begin{equation}
  a_{\rm preSN} = \frac{1}{1+e} \, a_{\rm postSN} . 
\label{eq21}
\end{equation}

Using (\ref{eq20}) and (\ref{eq21}) we find that the pre--SN
Roche lobe radius $R_R$ of the He star must obey 
\begin{equation}
R_R \simeq \frac{a_{\rm preSN}}{2} < 4.5\rsun \: \frac{1}{1+e} \: {\biggl[
         \frac{t_G}{\gamma} \frac{{(m_1+m_2)}^2 m_2^4}{m_1}
         \biggr]}^{1/5}  
\label{eq4}
\end{equation}
where we have assumed that $R_R$ occupies one--half of the pre--SN binary
separation

Finally, requirement (a) demands that the largest radius
$R_{\rm max}({\rm He})$ of the helium star must fit inside the pre--SN
binary. The evolution of the helium star is very rapid compared with
the orbital evolution so we assume that $a_{\rm preSN}$ is constant, and the
condition is $R_{\rm max}({\rm He}) < R_R$.
(If this
condition fails it is possible that some LMXBs nevertheless form after
a second 
common--envelope phase, cf Iben et al 1995b. We note, however, that
these authors assume an unusually efficient envelope ejection compared
to the standard common envelope description (\ref{eq17})
below. Application of the latter would almost always lead to a
merger. Hence we neglect this group of systems.)  
Iben et al.\ (1995b) fit $R_{\rm max}({\rm He})$ as
\begin{equation}
R_{\rm max} = 7700\rsun m_{\rm He}^{-5.5} \label{eq5}
\end{equation}
for $m_{\rm He}>2.5\msun$.

Hence combining (\ref{eq5}) with (\ref{eq4}) we see that (a) and (c) demand
\begin{equation}
{(m_1+m_2)}^2 m_2^4 > \frac{\gamma}{t_G} \: (1+e)^5 \, {
                      \biggl( \frac{3.87}{m_{\rm He}}
                         \biggr)}^{55/2} m_1 ,
                         \label{eq6} 
\end{equation}
thereby defining a lower limit for $m_2$ for given $m_1$, $m_{\rm He}$
(or a lower limit for $m_{\rm He}$ for given $m_1$, $m_2$). 

\subsubsection{Mass limits for NLMXBs}

Observational evidence strongly suggests that neutron stars emerge from the
supernova that creates them with a mass $m_1=1.4$. Figure 1 shows the
constraints (\ref{eq8}, \ref{eq6}) for this mass, with $t_G=1$ and
$\gamma=1$. (For
$m_2<0.3$, where magnetic braking is thought not to operate, (\ref{eq6})
was replaced by the corresponding constraint using $t_{\rm GR}$
instead of $t_{\rm MB}$.)
As can be seen
the two constraints cross at a value $m_2 =1.22, m_{\rm He} = 4.02$. Hence
$m_2 \geq 1.22$ for NLMXB formation through this channel.
%
We now go one step further and ask under
what conditions the binary can begin mass transfer with a secondary
which has spent less than a fraction $f$ ($0 < f \le 1$) of its lifetime
$t_{\rm MS}$ on the main sequence,  
i.e. with
\begin{equation}
t_{\rm MB} < f \: t_{\rm MS}. \label{eq9}
\end{equation}
Clearly this is a more stringent requirement than (c), where the rhs
is $t_{\rm Gal}> t_{\rm MS}$, and it therefore requires a larger mass $m_2$.
Taking
\begin{equation}
t_{\rm MS} = 10^{10}m_2^{-3}\ {\rm yr} \label{eq10}
\end{equation}
and following the same line of reasoning that led to the constraint
(\ref{eq6}), i.e.\ replacing $t_G$ in (\ref{eq6}) by $(f/m_2^3)$, we find
\begin{equation}
(m_1+m_2)^2 m_2 > \frac{\gamma}{f} \: (1+e)^{5}
                   {\biggl( \frac{3.87}{m_{\rm He}} \biggr)}^{55/2}
                   m_1, 
\label{eq12} 
\end{equation}
which must hold simultaneously with (\ref{eq6}). Condition (\ref{eq12}) is
best understood as a {\em lower} limit for the He star mass for any
given $m_1$, $m_2$.
In Fig.~1 we show the constraint for $f=1$ (dotted curve) and $f=0.25$
(dash-dotted), with standard magnetic braking ($\gamma=1$).
Instead of the rough approximation (\ref{eq10}) we used the 
fitting formulae
given by Kalogera \& Webbink (1996a) to determine $t_{\rm MS}$.
We see that condition (\ref{eq12}) with $f=1$ 
intersects the allowed mass range $1.2 \lta m_2 \lta 1.5$ set by
the Galactic age constraint and the requirement of thermally stable
mass transfer respectively.
Systems with $1.2 \lta m_2 \lta 1.5$ to the left of
the dotted line in Fig.~1 (hatched, wide spacing) turn--on mass
transfer with a giant donor at long orbital period
($P \gta 1 - 2$ d). As in most of them mass transfer is unstable 
(e.g.\ Kalogera \& Webbink 1996a) they would probably not appear as
long--period NLMXBs; we do not consider this group further. 
Systems with $1.3 \lta m_2 \lta 1.5$
to the right of this line (hatched, narrow spacing) will evolve to
short orbital periods. The location of the dash--dotted line in Fig.~1
clearly demonstrates that all secondaries of systems in
this group are somewhat evolved.
In other words: in the limit of negligible kick velocities 
secondaries in a large fraction of NLMXBs with
$P \lta 1 - 2$ d are close to the end of their MS lifetimes and are
partially nuclear--evolved ($\hat m_2 < 1$)
when mass transfer starts. The resulting low mass transfer rates (Pylyser \&
Savonije 1988b) then probably
explain why a comparatively large fraction of NLMXBs of both types
appear as transients. 
The simple scalings of Paper 1 suggest that SXT behaviour 
requires $\hat m_2 \lta 0.25$; full population 
synthesis calculations are needed in order to quantify the resulting SXT 
fraction. 

We note that Kalogera \& Webbink (1996b) find that short--period LMXBs
do not form at all without kick velocities. 
The reason for this difference from our conclusions is 
a modification of the standard form (\ref{eq1}) of magnetic braking
such that $t_{\rm MB}$ varies like $m_2^{-2}$
(rather than $m_2^{-4}$), and the introduction of a reduction factor
for 
secondaries more massive than $1\msun$. Even if the reduction factor
is ignored (which is $\simeq 7$ for $m_2=1.5$), their $t_{\rm MB}$ is
still longer by a factor $\simeq 6$ for  
$m_2=1.5$. This corresponds to $\gamma = 6$ in (\ref{eq12}), or,
equivalently, to $f=1/6$.
Inspection of Fig.~1 shows that the critical line corresponding to
$f=1/6$ would be to the right of the dash--dotted $f=1/4$ line, hence
no parameter space would be left for short--period NLMBX formation. 
Physically this means that due to the correspondingly slower evolution
the systems would not reach contact within the age of the
Galaxy. 

This underlines the sensitivity of our result to the strength
of magnetic braking for main--sequence stars in the critical mass
range where the convective envelope disappears. To illustrate this
further we consider the lower limit $a_{\rm crit}$ we obtain 
for the post--SN orbital distance $a_{\rm postSN} \simeq 2 (1+e) R_R$
when the requirement $R_R > R_{\rm max}$(He) is combined with 
(\ref{eq5}) and (\ref{eq8}):
\begin{equation}
   a_{\rm postSN} > a_{\rm crit} = 5.1 \rsun \: (1+e) \: {\left(
                             \frac{4.3}{m_2 + 2m_1}\right)}^{5.5}  .
\label{eq12a}
\end{equation}
Note that this limit does not depend on magnetic braking and that 
$a_{\rm crit}$ is essentially determined by $m_2$ ($m_1=1.4$, $0 \leq
e \leq 1$). For NLMXBs with main--sequence donors 
of mass $m_2$ to exist at all the detached evolution time $t_{\rm MB}$
needed to shrink the orbit from $a_{\rm crit}$ into contact 
must be shorter than $t_{\rm MS}$. On the other hand,
to ensure that a large fraction of the secondaries are significantly
evolved when mass transfer turns on, $t_{\rm MB}$ must be longer than
a certain minimum fraction $f\simeq 1/4$ of $t_{\rm MS}$. This places
a tight limit on the allowed strength of magnetic braking for systems
with secondary mass $m_2$, initial orbital distance $a_{\rm crit}$ and
neutron star mass $m_1=1.4$:
\begin{equation}
  0.25 \la \frac{t_{\rm MB}(a_{\rm crit}, M_2)}{t_{\rm MS}} \leq 1 ,
\label{eq12b}
\end{equation}
independent of the functional dependence of $t_{\rm MB}$ on mass and
period.

\subsubsection{Mass limits for progenitor systems}

The narrow range 
\begin{equation}
1.3 \lta m_2 \lta 1.5 \label{eq13}
\end{equation}
also implies a very restricted range for 
$m_{\rm He}$: from Fig.~1 we see that this must obey
\begin{equation}
4.0 \lta m_{\rm He} \lta 4.3. \label{eq14}
\end{equation}
Using the analytic fit
\begin{equation}
\log m_{\rm He} = 1.273 \log m_{\rm prog} - 0.979 \label{eq15}
\end{equation}
from Kalogera \& Webbink (1996b) to stellar models by Schaller et al.\
(1992) we see that the mass of the MS
progenitor of the helium (and ultimately neutron) star must be in the range 
\begin{equation}
17.5 \lta m_{\rm prog} \lta 18.5. \label{eq16}
\end{equation}
Immediately before the CE phase the primary has evolved beyond core He
exhaustion and has lost mass in the form of stellar
winds. Interpolation of Schaller et al.'s models suggest that
primaries from the mass range (\ref{eq16}) will then have masses
$m_{\rm preCE}$ in
the range 
\begin{equation}
15.2 \lta m_{\rm preCE} \lta 15.9. \label{eq16a}
\end{equation}
Wind mass loss carrying the specific 
a.m.\ of the mass--losing component increases the orbital
separation as the inverse of the total mass. Hence the MS parent
binary orbit is closer 
by a factor $\simeq 0.9$ than the immediate pre--CE binary orbit.

Simple treatments
of common--envelope evolution (e.g.\ WK) lead to the relation
\begin{equation}
a_{\rm preCE} \simeq {2m_{\rm preCE}(m_{\rm preCE}-m_{\rm He}+\alpha \,
       \lambda r_R \, m_2/2)
      \over \alpha \, \lambda  r_R \, m_{\rm He}m_2} a_{\rm postCE},
     \label{eq17}
\end{equation}
where $a_{\rm preCE}, a_{\rm postCE}$ are the binary separation before and
after the common--envelope phase, $\alpha$ is the fraction of the orbital
binding energy used to drive off the envelope, $\lambda$ is a weighting
factor for the gravitational binding energy of the envelope to the core,
and $r_R = r_R(m_{\rm preCE}, m_2)$
is the fractional size of the primary's Roche lobe at the start of the CE 
phase. 
We find with the mass limits (\ref{eq13}, \ref{eq14}, \ref{eq16a}) 
and $r_R \simeq 1/2$, $\lambda \simeq 1/2$
\begin{equation}
a_{\rm preCE} \simeq \frac{8}{\alpha} \frac{m_{\rm preCE}}{m_2}
\frac{m_{\rm preCE} - m_{\rm He}}{m_{\rm He}} 
\simeq {250 \over \alpha} a_{\rm postCE}. \label{eq18}
\end{equation}
Since there is little orbital evolution before the helium star undergoes its
supernova explosion, and the post--SN separation is within a factor
of 2 of the pre--SN separation, we can translate this into limits on
$a_{\rm preCE}$, or equivalently the orbital period of the progenitor binary.
We get an upper limit from requirement (c) (cf \ref{eq4}), and a lower limit
from requirement (a), using (\ref{eq5}) and the fact that $m_{\rm He} < 4.3$.
(This limit prevents NLMXB progenitors emerging from the 
common--envelope phase with the MS secondary already close to filling its Roche
lobe, which is WK's condition 6; the requirement here gives a tighter limit.
This in turn justifies our assumption above that the post--SN period $P$
was much larger than the contact period $P_c$.)
The resulting 
constraints are close, i.e. $\sim 2300 \rsun/\alpha$ and 
$\sim 1300\rsun/\alpha$ respectively. Together they
show that the orbital distance $a_{\rm prog}$ and period $P_{\rm
prog}$ of the MS progenitor $\sim 18\msun + 1.4\msun$ binary must lie
in the range  
\begin{equation}
1200 \ \rsun \lta a_{\rm prog}/\alpha \lta 2100 \ \rsun
\label{eq19a}
\end{equation}
and 
\begin{equation}
3 \ {\rm yr} \lta \alpha^{3/2} P_{\rm prog} 
            \lta 7\ {\rm yr}. \label{eq19}
\end{equation}
The formation channel we have considered requires that the progenitor
fills its Roche lobe after core--heluim exhaustion but before it
would explode as a SN. For a $18\msun$ MS star this implies $1200 
\la a_{\rm prog}/\rsun \la 1600$ or $3 \la P_{\rm prog}/{\rm yr} \la 4.6$ 
(Kalogera \& Webbink 1996b), thus reducing the limits (\ref{eq19a}),
(\ref{eq19}) even further. 
If  most short--period NLMXBs in the Galaxy form via helium--star
supernovae, this restricted range of allowed inital orbits and 
the 
very narrow limits (\ref{eq13}, \ref{eq16})
then show that only an extremely small progenitor population can
produce these systems, explaining their
comparative rarity. 
Following Iben et al.\ (1995b) a very
crude estimate of the Galactic short--period NLMXB formation rate
$\nu$ can be obtained from 
\begin{equation}
  \nu = 0.2 \, \Delta \log a_{\rm prog} \, \frac{\Delta m_{\rm
  prog}}{m_{\rm prog}^{2.5}} \frac{\Delta m_2}{m_{\rm prog}}\  {\rm yr}^{-1} ,
  \label{eq191}
\end{equation}
where $\Delta x$ denotes the allowed range of the quantity $x$.
Eq.~(\ref{eq191}) relies on standard assumptions for the distribution
of initial orbital parameters in newly forming ZAMS binaries, in
particular on a flat distribution in $\log M_2/ M_{\rm prog}$. 
Using the limits obtained above we find $\nu \simeq 2 \times
10^{-7} {\rm yr}^{-1}$. Replacing (\ref{eq191}) by 
the initial parameter distribution of ZAMS binaries used 
in Kalogera \& Webbink (1996b) gives a similar value for $\nu$.

\subsection{The role of kick velocities}

In reality supernova explosions are probably not
spherically symmetrical, and give the compact primary remnant a kick
velocity. In this case
the main differences from the considerations in 2.1 are twofold
(see e.g.\ Kalogera 1996a).
First, a suitably directed kick can
unbind binaries that would have been stable according to (\ref{eq8}), or
keep together binaries that would have disrupted according to this
inequality. The latter case
requires a kick velocity comparable to the pre--SN orbital
velocity of the companion, directed almost parallel 
to its instantaneous motion, and is therefore rather rare unless the
kick velocity is close to an optimum value near the pre--SN orbital
velocity.  
Second, relation (\ref{eq21}) no longer holds, and $a_{\rm postSN}$
can be either smaller or larger than $a_{\rm preSN}$ (but never larger
than $2 a_{\rm preSN}$).
These two effects not only widen the allowed region of NLMXB formation
in the $M_2 - M_{\rm He}$ plane of Fig.~1, but in particular allow
the formation of NLMXBs with almost unevolved secondaries.

In the absence of a theoretical understanding of the origin, magnitude and
direction of kick velocities the standard approach in population
synthesis considerations (e.g.\ Brandt \& Podsiadlowski 1995, Terman
et al.\ 1996, Webbink \& Kalogera 1996b)
is to assume that the kicks are isotropic and 
that their magnitude derives from a given distribution function,
characterized by a certain r.m.s.\ value $\sigma_{\rm k}$. 
As demonstrated e.g.\ extensively by Kalogera (1996a) the
resulting probability distributions can be expressed in
terms of the two governing dimensionless parameters,
the ratio $\xi$ of kick velocity to relative
orbital velocity in the pre--SN orbit, $\xi = \sigma_{\rm
k}/v_{\rm orbit}$, and the ratio $\beta$ of post--SN and
pre--SN binary mass, $\beta = (m_1 + m_2)/(m_{\rm He} + m_2)$.

Given the stochastic nature of the problem only a full population
synthesis can provide a quantitative estimate of the fraction of
NLMXBs with significantly nuclear--evolved secondaries at turn--on of
mass transfer. 
To illustrate the main effect of kick velocities
and to gain a very rough estimate of the maximum kicks we can tolerate
and still
maintain the large fraction of systems with evolved secondaries found
for spherically symmetrical SNe, we make use of the analytic expression
for the distribution of binaries over $\alpha_c = a_{\rm
postSN}/a_{\rm preSN}$ derived by Kalogera (1996a) under the 
assumption of a Maxwellian kick velocity distribution.
In Fig.~2 we show the same limits as plotted in Fig.~1, but with the
factor 
$(1+e)$ in (\ref{eq6}) and (\ref{eq12}) replaced by the approximate
median value of $\alpha_c$, $1.75$, $1.25$ and $0.75$, for small ($\xi
= 0.1$), moderate ($\xi = 0.3$) and large kick velocities ($\xi =
1.0$), respectively. These median values  depend only
weakly on $\beta$. For strong kicks the critical line (\ref{eq8}) ---
which is equivalent to $\beta = 0.5$ --- was replaced by the line
$\beta = 0.4$ as the survival probability of these systems is only
a factor of 2 smaller than the one for the most stably bound systems
(corresponding to $\beta \simeq 0.75$). 
The resulting enclosed area in the $M_2 - M_{\rm He}$ plane can be
thought of as representative for the effective parameter space of
NLMXB formation. 
Assuming that the area is a measure of 
the corresponding relative formation rate, 
Fig.~2 suggests that in the case of standard magnetic braking
(\ref{eq1})
systems with unevolved secondaries still constitute only a small
fraction of NLMXBs 
for $\xi = 0.1$,
represent the majority for $\xi = 0.3$, and entirely dominate for $\xi
= 1.0$. 

More efficient magnetic braking would increase the dominance of
unevolved systems even further, whereas a favourable combination of
weaker magnetic braking and a kick velocity distribution with small
(or moderate) mean velocity could both ensure the formation of the
(otherwise forbidden) class of short--period NLMXBs, and the
predominance of nuclear-evolved main--sequence donors. However, a
population subject to a large mean kick velocity would necessarily
contain a large fraction of systems where the secondary is close to
contact in the post--SN orbit and hence essentially unevolved at
mass transfer turn--on --- whatever the strength of magnetic
braking. The reason for this is twofold. First, the limit
(\ref{eq12a}) formally allows post--SN orbital periods shorter than
the contact period $P_c$ for $m_{\rm He} \ga 4.7$, and with strong
kicks a large fraction of such binaries would survive the SN.
Second, the SN--induced orbital reduction factor $\alpha_c$ is very
small for the majority of systems.

In view of this, the large fraction of soft
X-ray transients observed among NLMXBs provides a strong argument
not only against a magnetic braking stronger than our standard case
(\ref{eq1}), but also  
against a mean kick velocity of order $350 - 400$~kms$^{-1}$ ($\xi
\simeq 1$) invoked by Lyne \& Lorimer (1994) from 
observed pulsar proper motions. Fig.~2 shows that kick
velocities must on average be small compared to the pre--SN orbital
velocity, probably $\xi \la 0.1$. 
This is consistent with a more 
recent reevaluation of the initial velocity distribution of radio
pulsars which confirms the existence of the high--velocity tail found
by Lyne \& Lorimer, but   
suggests that the distribution has its maximum at zero velocity, hence
a smaller average value (Hansen \& Phinney 1996, 
Hartmann 1996).

\subsection{Alternative evolutionary channels}

Short--period NLMXBs might also  
form from systems with initially fairly massive
main--sequence secondaries above the limit for thermally stable mass
transfer, $1.5 - 2 \lta m_2 \lta 3$, provided these
survive the initial phase of thermal timescale (hence
super--Eddington) mass transfer. (Systems with $m_2 \ga 3$ would
probably develop a delayed dynamical instability, see Hjellming 1989).
Kalogera \& Webbink (1996a, b) pointed out that such systems could
reappear as stable NLMXBs with sub--Eddington mass 
transfer rate and donor mass $1 \lta m_2 \lta 1.5$.
An analysis similar to the one in 2.1.2 (using $t_{\rm GR}$ instead of
$t_{\rm MB}$) reveals that a significant fraction of them would emerge
from the CE phase almost semi--detached. 
They would turn on mass transfer with an unevolved secondary, again in
conflict with the observed comparatively large fraction of transients
among NLMXBs.  
Hence we conclude that this channel cannot contribute significantly to the
formation of short--period NLMBXs.

Recently, Kalogera (1996b) has described yet another formation
mechanism for 
LMXBs where no CE phase is involved. Instead the orbital shrinkage is
achieved by a suitably directed kick when the primary star in the
wide progenitor binary explodes as a SN before it reaches its Roche
lobe. Population synthesis models show (Kalogera 1996b) that the
production of short--period LMXBs via this channel is altogether
negligible compared 
to the standard He--star SN case if the kick velocities are large
($\sigma_{\rm k}$ of order $300 - 400$~kms$^{-1}$), but 
might account for a formation rate comparable to the one derived in
2.1.2 if $\sigma_{\rm k}$ is close to an 
optimum value $\simeq 50$~kms$^{-1}$. However, in the latter case a 
relatively large fraction of LMXBs would start the X-ray phase 
with a very small (below $0.3 \msun$) donor mass, hence with
essentially unevolved secondaries. Again, this suggests that the direct SN
mechanism represents only a minor channel for the formation of
short--period NLMXBs (whereas it might be important to produce 
long--period systems with $P_d \gta 100$ d) even if kick velocities are
generally small.

\section{NLMXBs BELOW THE PERIOD GAP}

The difference between the period histograms of LMXBs and cataclysmic variables
(CVs), in which the primary is a white dwarf, has long been remarked
(e.g.\ White \& Mason 1985, van~den~Heuvel \& van~Paradijs 1988, Verbunt \&
van~den~Heuvel 1995, Kolb 1996). 
The most prominent difference is a total (or near--total) lack of
LMXBs in the 
$80 < P < 120$ min region below the famous CV period gap.
A KS test reveals that there are also statistically significant
differences above the gap. The
hypothesis that the LMXB and CV samples 
are drawn from the same underlying distribution can be rejected at a 
confidence level
$>99.99\%$ in the period range $80$~mins -- 2~d (Fig.~3a), and 
at $\simeq 99.96\%$ in the range 3~h -- 2~d (Fig.~3b).
This cannot be explained by selection effects discriminating against 
short orbital periods,
since many of the X--ray periods were turned up by satellites such as EXOSAT
which had a 4--day orbit. 
The differences conflict with the simple picture 
of CVs and LMXBs as essentially the same in terms of their secular evolution,
apart from the substitution of a white dwarf by a neutron--star or black--hole
primary. However our arguments above show that this simple picture is 
inaccurate, particularly for NLMXBs. CVs emerge from common--envelope
evolution with a full range of secondary masses down to $m_2 \sim 0.1$. The 
vast majority of these stars are essentially unevolved, and the post--CE 
separations are so small in many cases that the secondaries are close to 
their Roche lobes. The majority of CVs probably start mass transfer
at periods below the gap
(more than 67\% according to King et al., 1994). None of these features
hold for NLMXBs, as we have seen: the secondaries are confined to the narrow 
range $1.3 \lta m_2 \lta 1.5$ initially, a large fraction of them must
be significantly  
nuclear--evolved, the post CE (and SN) Roche lobes are considerably larger than
the secondaries, and they start mass transfer at periods in the range
$10$~hr $\lta P \lta 30$~hr. 

We investigate differences between the CV and NLMXB period histogram 
which arise alone from these effects in Fig.~4.
In particular, we test if the population of NLMXBs below $P \sim 2$ hr
is much smaller than for CVs, as the 
lack of the enormous influx of newly--formed systems boosting
the CV distribution there would suggest. 
Despite this lack the expected intrinsic period distribution (Fig.~4,
middle panel) derived from a typical evolutionary sequence (Fig.~4,
upper panel) still predominantly populates the short period range,
simply because the period evolves more slowly there (number density
$\propto \dot P^{-1}$).  
However, these short--period systems are suppressed 
for any visibility function $\propto (-\dot M_2)^\alpha$ 
with $\alpha \ga 1$ (Fig.~4, lower panel). 
Such a visibility function corresponds for example to a
flux--limited sample taken from a disc--like population.
In addition, the detection probability function obtained
in this way shows a pronounced peak at long orbital periods (here
$P\simeq 8$~hr), a feature consistent with the observed LMXB
overpopulation 
at long orbital periods compared to CVs (see Fig.~3b).
This peak is a consequence of the high mass transfer rate immediately
after contact is reached in NLMXBs, since the systems
are close to instability ($M_2 \sim M_1$). 
The complexity of quantifying the relevant selection effects makes it
difficult to decide if $\alpha \ga 1$ properly describes the NLMXB
population. However, only if indeed $\alpha < 1$ is an 
additional mechanism needed to account for the lack of 
short--period systems. One such mechanism is the evaporation of the 
secondary star by pulsar irradiation from a rapidly rotating neutron
star spun--up by accretion (van~den~Heuvel \& van~Paradijs 1988). 
Remarkably, this implicitly assumes that all LMXBs cross the gap,
i.e.\ assumes the result we have demonstrated above.

\section{CONCLUSIONS}

We have shown that short--period neutron star low--mass X-ray binaries
forming from  
helium--star supernovae without kick velocities must have
secondaries with masses in the narrow range $1.3\msun \lta M_2 \lta
1.5\msun$. 
For standard strenghth magnetic braking the secondary star
is already  
significantly nuclear--evolved when mass transfer begins, 
explaining why the resulting mass transfer rates are in many cases
low enough for a substantial fraction of these systems to appear
as soft X--ray transients even at short ($P \lta 1 - 2$~d) orbital periods
(cf Paper 1). 
On the other hand, if the neutron star receives a strong kick velocity
at birth, many NLMXBs would form with unevolved low--mass donors.  
The observed large fraction of SXTs among NLMXBs then forces us to 
conclude that kick velocities must be small compared to the
pre--SN orbital velocity ($\la 10\%$, i.e.\ $\la 50$~km/s) for a large
fraction of progenitor systems. This is consistent with a recent
reevaluation of observed pulsar proper motions which suggest that 
the distribution of neutron star velocities at birth has a maximum at
zero. 
Similarly, short--period NLMXBs forming from both initially
thermally unstable systems (Kalogera \& Webbink 1996a, b) and 
via the direct SN channel (Kalogera 1996b) would have a large fraction
of (low--mass) unevolved secondaries, suggesting that neither of these
channels contribute significantly to the short--period NLMXB
population.

Ignoring the uncertain survivors of thermally unstable mass transfer,
the very special formation conditions for the case with negligible
kick velocity also show that only a very restricted progenitor
population (essentially $18\msun + 1.4\msun$
binaries with periods $P_{\rm prog} \sim 4$ yr) can form NLMXBs,
explaining their rarity in the Galaxy. We estimate a total formation
rate $\sim 2\times 
10^{-7}$ yr$^{-1}$. The formation conditions are much more 
restricted than for cataclysmic variables. In particular short--period NLMXBs 
must all begin mass transfer at periods $\gta 12$ hr, in contrast to CVs,
of which a majority start mass transfer at periods $\lta 1 - 2$
hr. The resulting 
NLMXB period histogram has far fewer systems at short periods than the CV 
version, in agreement with observation.

The smallness of the area of the $M_{\rm He} - M_2$ plane (Fig.~1) allowing
short--period
NLMXB formation is very striking. The fact that the resulting population has
several properties in good agreement with observation is implicit confirmation
that the assumed formation conditions are realistic. In particular it is clear
that the neutron star mass at formation cannot be significantly larger than
$1.4 \msun$: if it were, 
the allowed area would become much larger, sharply decreasing the 
predicted relative population of transient systems. However, we can say nothing
about the lower limit on the formation mass, as systems with $m_1 \lta 1.2$
would not appear as NLMXBs (the allowed area in Fig.~1 would
disappear). This result suggests another conclusion:  
with initial conditions $1.2 \lta m_1 \lta 1.4, 1.2 \lta m_2 \lta 1.5$, 
evolution without mass loss would lead us to expect neutron star masses 
$\gta 2.4\msun$ in short--period systems, or those where direct estimates of 
$m_2$ give a low mass, such as Cen X--4 (Shabaz et al.\ 1993). 
As there is no observational support for such masses, 
this suggests that a large fraction 
of the mass transferred in the NLMXB phase is lost from the binary. The most
likely way for this to occur is through mass loss from the accretion discs
in these systems (cf Begelman, McKee \& Shields, 1983; Czerny \& King, 
1989).
If the mass is lost at disc radii much 
greater than the size of the neutron star the central accretion rate can in
principle be smaller than the mass transfer rate, so it may be simplistic
to infer the latter from the observed X--ray flux. This in turn would mean
that the presence of transient behaviour would pose a somewhat less stringent
upper limit on the mass transfer rate than we inferred in Paper 1.

This paper has discussed the formation of LMXBs containing neutron stars. 
The constraints on the formation of black--hole LMXBs appear to be
weaker (cf, e.g., Romani 1994, 1996),
as are the conditions for them to appear as transients (Paper 1). We shall 
investigate this problem in a future paper.

\acknowledgements
We thank Ron Webbink and Vicky Kalogera for providing us with a copy
of their papers on LMXB formation prior to publication and for very
helpful discussions. We also thank the referee for useful comments.
This work was supported
by the UK Particle Physics and Astronomy Research Council through a Senior
Fellowship (ARK) and a Rolling Grant for theoretical astrophysics
to the Leicester Astronomy Group.

\clearpage


\begin{figure}
\vspace*{-2.5cm}
\plotone{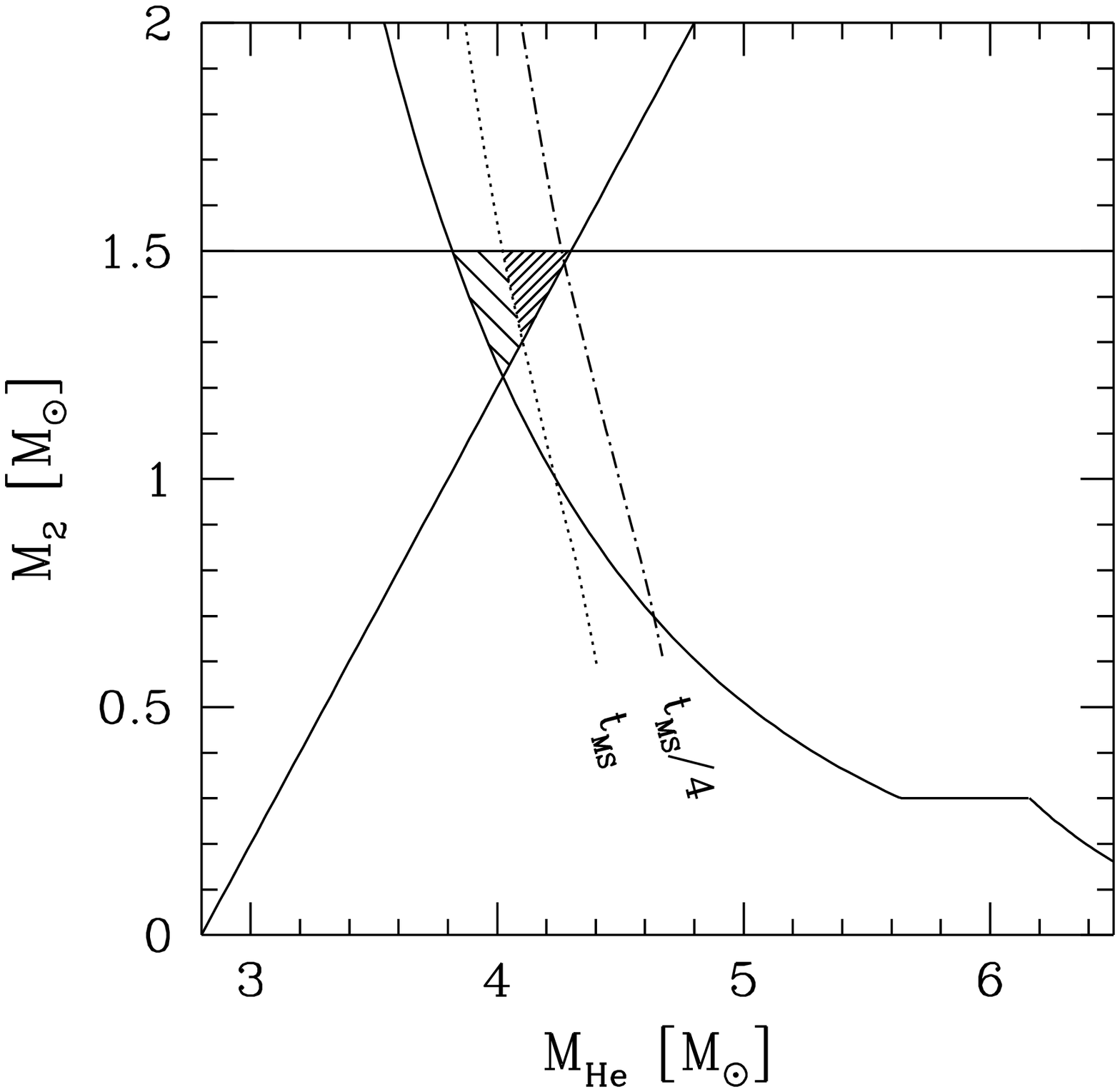}
\caption{ {\small
%
Constraints on NLMXB formation in the donor mass ($M_2$) -- He star
mass ($M_{\rm He}$) plane in the absence of kick velocities. 
The straight line with positive slope
represents the SN survival condition (\ref{eq8}) and is a lower limit
for $M_2$. The curve with negative slope represents another lower  
limit for $M_2$ (or, equivalently, $M_{\rm He}$), the Galactic age constraint
(\ref{eq6}) (for 
$M_2>0.3\msun$; as magnetic braking does not operate in systems with 
fully convective secondaries we plot for $M_2<0.3\msun$ the corresponding 
limit obtained by replacing $t_{\rm MB}$ by $t_{\rm GR}$).
The horizontal line represents an upper (stability) limit for
$M_2$ (where $1.4\msun$ was assumed for the neutron star mass).
These limits restrict the allowed range of NLMXB formation to 
the small triangular (shaded) area enclosed by these curves.
Also shown is the lower limit 
for $M_2$ (or $M_{\rm He}$) if 
the system's age is less than $t_{\rm MS}$ (dotted curve; $f=1$
in eq~[\ref{eq12}]), or less than $0.25 t_{\rm MS}$ (dash--dotted;
$f=0.25$) at mass transfer turn--on. The allowed range for
short--period NLMXBs with main--sequence donors is only the
narrow--hatched area. 
}}
\end{figure}

\clearpage

\begin{figure}
\vspace*{-2.0cm}
\plotone{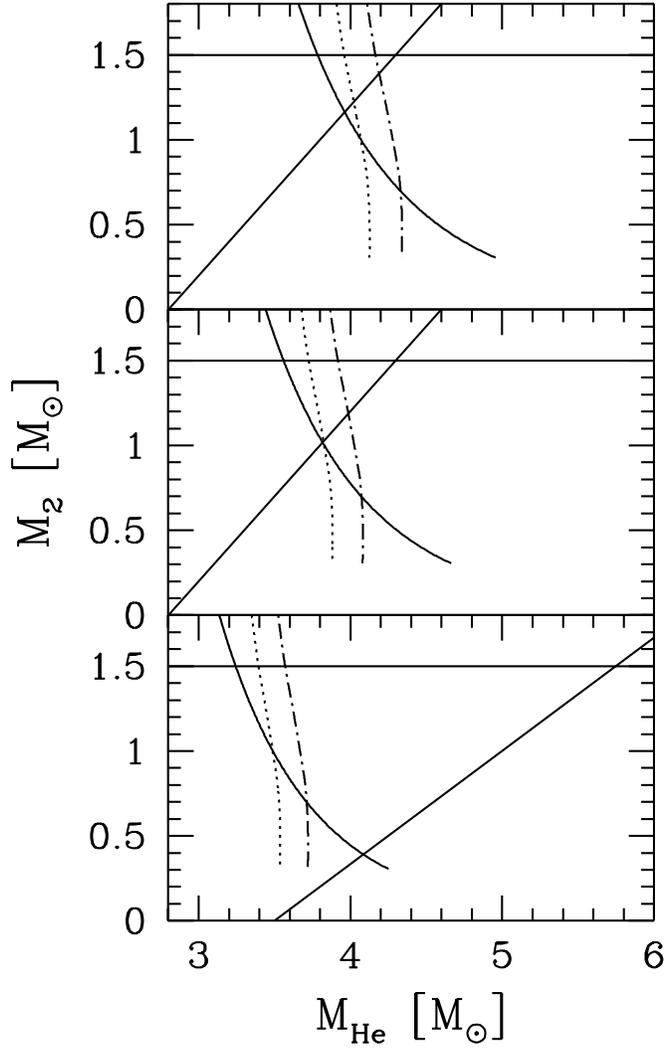}
\caption{ {\small
Same as Fig.~1, but for weak ($\xi=0.1$, upper panel), moderate ($\xi
= 0.3$, middle panel) and
strong ($\xi=1.0$, lower panel) kick velocities. For the $\xi=1$ case
the SN survival line was replaced by a line with $\beta={\rm
const.}=0.4$ (see Sect.~2.2). 
}}
\end{figure}

\begin{figure}
\vspace*{-2.5cm}
\plotone{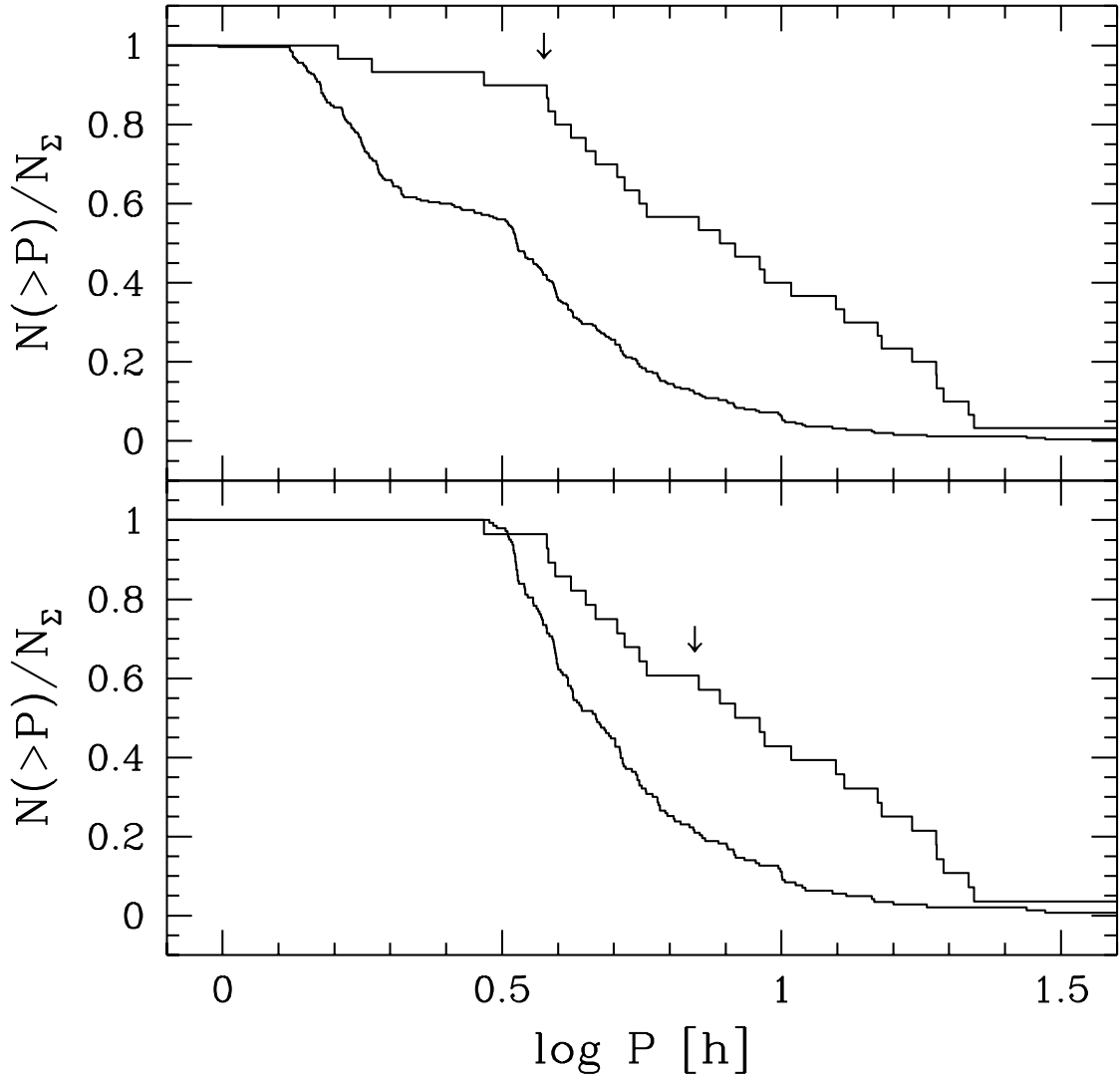}
\vspace{1cm}
\caption{ {\small
Normalized cumulative period distribution for CVs and LMXBs in
selected period ranges, data taken from Ritter \& Kolb (1996). 
{\em a: (upper panel)} systems with orbital periods between 80 min and
2~d (249 CVs, 30 LMXBs); 
{\em b: (lower panel)} systems with orbital periods between $\simeq3$~h and
2~d (142 CVs, 27 LMXBs).
The arrows indicate the period where the difference between the
distributions is largest.
}}
\end{figure}

\begin{figure}
\vspace*{-1.0cm}
\plotone{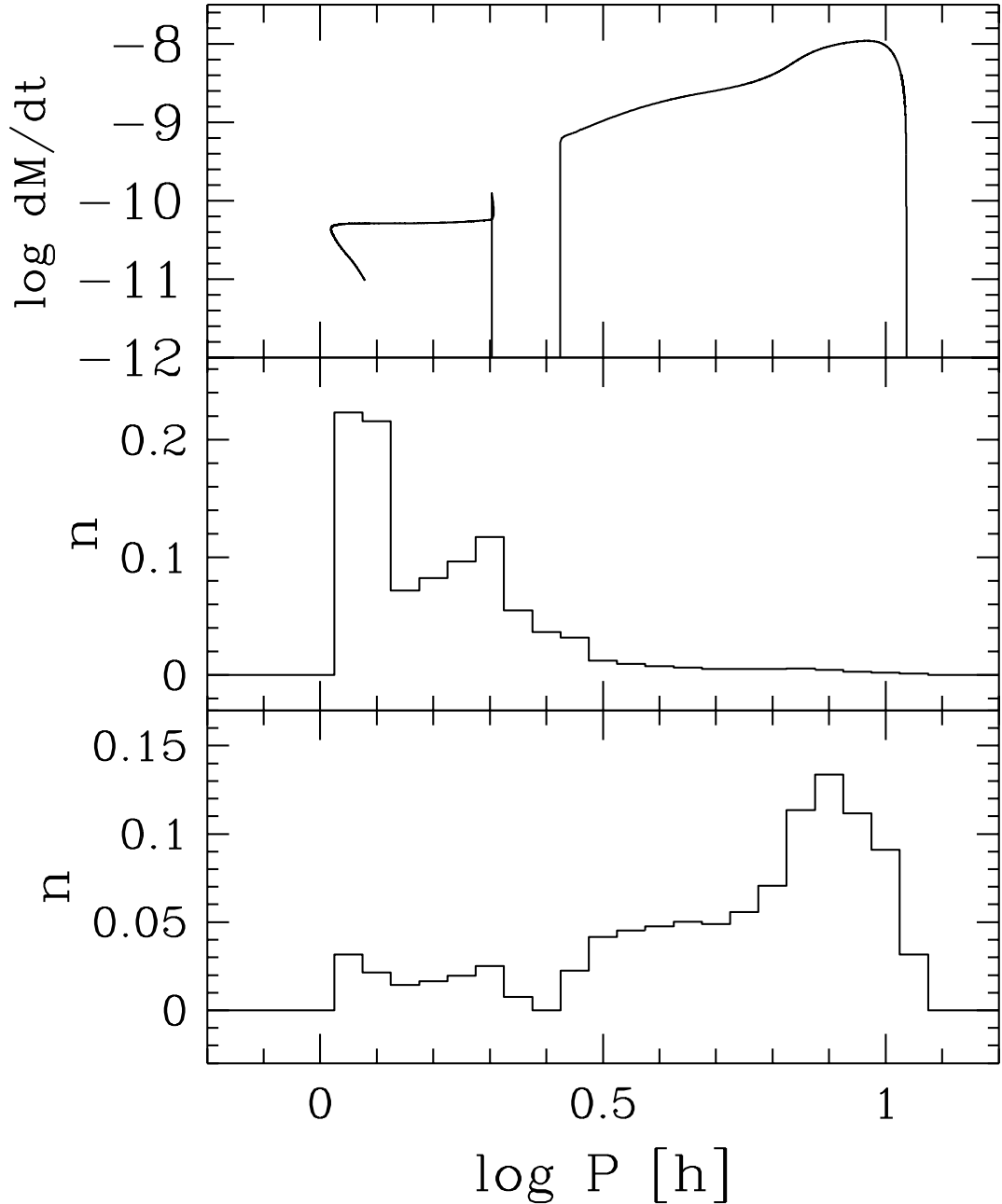}
\vspace{-1.3cm}
\caption{ {\small
Possible shape of the NLMXB period distribution if all systems
form with a secondary which is somewhat nuclear--evolved and has a rather high
mass. As a representatitve evolutionary sequence we use a calculation
by Singer (Singer et al.\ 1993, Ritter 1994) with an initial donor
mass $M_2=1.2\msun$, initial central hydrogen mass fraction 0.36
(i.e.\ with age $= 0.5 t_{\rm MS}$ at turn--on),
and a (constant) mass $M_1=1.0\msun$ for the compact object. The mass
transfer rate $\dot M$ as a function of orbital period $P$ for this
sequence is shown in the upper panel. The sequence was originally
used to represent CV evolution; a slightly different initial primary
mass ($1.4 \msun$ instead of $1.0\msun$) does not affect the main
conclusions drawn in the text.
The middle panel depicts the intrinsic discovery probability
$\propto P/\dot P$ vs $\log P$, a quantity representing the intrinsic
period distribution of a population of such binaries with similar
initial configurations at mass transfer turn--on.
In the lower panel we plot the corresponding discovery probability
obtained by multiplying by a visibility factor $\propto \dot M$ to
account for observational selection. 
}}
\end{figure}

\end{document}